\documentclass[conference, twocolumn]{IEEEtran}
\IEEEoverridecommandlockouts
\usepackage{cite}
\usepackage{amsmath,amssymb,amsfonts,bm}
\usepackage{graphicx} 
\usepackage{textcomp}
\usepackage{algcompatible}
\usepackage{algorithm}
\usepackage{algpseudocode}
\algdef{SE}[DOWHILE]{Do}{doWhile}{\algorithmicdo}[1]{\algorithmicwhile\ #1}%

\newcommand{\Break}{\State \textbf{break} }
\usepackage{float} 

\usepackage{blindtext}
\usepackage{eso-pic}

\makeatletter

\def\ps@IEEEtitlepagestyle{%
	\def\@oddfoot{\mycopyrightnotice}%
	\def\@evenfoot{}%
}
\def\mycopyrightnotice{%
	{\footnotesize  978-1-7281-1508-5/19/ \$31.00 \textcopyright 2019 IEEE\hfill}
	{}
	\gdef\mycopyrightnotice{}
}

\usepackage{eso-pic}
\newcommand\AtPageUpperMyright[2]{\AtPageUpperLeft{%
		\put(\LenToUnit{0.08\paperwidth},\LenToUnit{-1.1cm}){%
			\parbox{0.5\textwidth}{\raggedleft\fontsize{10}{11}\selectfont #1}}%
	}}%
	\newcommand{\conf}[1]{%
		\AddToShipoutPictureBG*{%
			\AtPageUpperMyright{#1}
		}
	}

\begin{document}
\conf{27th Iranian Conference on Electrical Engineering (ICEE2019)}
\title{Improving Energy Efficiency of Massive MIMO Relay Systems using Power Bisection Allocation for Cell-Edge Users}

\author{Mohammad Koolivand$^\dagger$, Mohammad Hossein Bahonar$^\dagger$, Mohammad Sadegh Fazel$^\dagger$\\
$^\dagger$ Deptartment of Electrical and Computer Engineering, Isfahan University of Technology, Isfahan, Iran 84156-83111\\
Emails: \{kulivandmohamad@gmail.com, mh.bahonar@ec.iut.ac.ir, fazel@cc.iut.ac.ir\}
}
\maketitle

\begin{abstract}
This paper studies downlink energy efficiency (EE) of a cellular massive Multiple Input Multiple Output (MIMO) system with massive number of antennas at both base station (BS) and relay considering cell-edge users which is less investigated in the literature.
Some improving methods for base station (BS) and relay power allocation under quality of service constraints are proposed.
The EE optimization problem is provided over BS and relay power variables (power dimensions).
Considering the quasi-concavity property of the EE function relative to each power variable, we propose to use a power bisection algorithm (PBA) in one dimension, followed by the exhaustive search in the other dimension (ODS algorithm) which is called PB-ODS algorithm and also a method to limit the range of power variables.
The results show that the performance of the PB-ODS algorithm approaches the performance of optimal solution which is exhaustive search in both dimensions while it has a lower complexity.
In addition to that we suggest to use the PBA in two dimensions alternatively as the sub-optimal alternative optimization (AOP) algorithm. 
The complexity can highly be reduced using the AOP algorithm with a slight but acceptable degradation in performance which makes the algorithm much suitable for practical use.
\end{abstract}

\IEEEpeerreviewmaketitle

\section{Introduction}\label{Intro}
Due to the ever increasing rate demand in cellular netwroks, new technologies and architectures such as heterogeneous networks\cite{HetNet}, device-to-device communication\cite{MHB_2018}, and Massive MIMO\cite{c3} have been proposed in order to increase the overall sum-rate or spectral efficiency (SE) of wireless systems.
Massive MIMO along with cooperative relaying systems, which is known as one of the most important technologies for increasing network capacity and network coverage, can greatly help to meet some future network's requirements and can be considered as a good choice for the next generation of wireless communication systems. In the early researches\cite{wang2005capacity, bolcskei2006capacity}
, some challenges such as system capacity have been addressed.

In \cite{c4}, the sum rate of a system where several pairs of single-antenna users communicate with each other through a one-way amplify and forward (AF) MIMO relay is obtained. In \cite{c10, c11}, considering a two-way AF MIMO relay system, maximizing the total network rate is investigated. 
In \cite{c12} a full-duplex model for a one-way AF relay system with maximum ratio combining/transmission (MRC/MRT) is proposed and a lower bound for the total transmission rate of the network is obtained using Jensen inequality. 
The scenario considered in \cite{c3} is a one-way massive MIMO AF relay system in which its energy efficiency (EE) is investigated and some methods based on Random Matrix Theory are proposed to improve it. 
In \cite{gayan01}, the achievable spectral efficiency (SE) of a multi-cell AF multi-way massive MIMO relay network in presence of some practical transmission impairments, including imperfect channel estimation or outdated channel state information (CSI) has been investigated.

In recent years, some researchers have been conducting investigations on massive MIMO relay systems in cellular networks. 
Researchers in \cite{c13} have considered a cellular network with MIMO relay and investigated its total SE. 
Different configuretions for single-cell massive MIMO relay systems have been performed recently in \cite{2018massive}.
In \cite{huang2013energy}, authors investigate the tradeoff between SE and EE in AF relay networks.
The system model of \cite{c14} includes a large number of single-antenna users, which communicate with a massive MIMO base station (BS) with aid of single-antenna AF relays. 
The EE and SE of the system is obtained using the bisection algorithm and is compared with the exhaustive search method. 

Due to increasing power consumption in recent years, design of energy efficient systems have attracted a lot of researches.
Because of the importance of EE in massive MIMO systems, we have investigated the combination of massive MIMO and cooperative relaying systems with energy efficient structures for 5G and beyond cellular networks. 
Few papers have investigated the EE in a cellular massive MIMO relay-aided network while the SE and the capacity of the system is usually examined; for example \cite{myAccess}  investigates massive MIMO two-way relays in cellular networks.
So in this paper we study the EE of a single-cell massive MIMO relay system and model the power allocation problem using a non-convex optimization problem.
We aim to serve cell-edge users in a downlink transmission scenario.
Circuit power consumption is considered in the power consumption model.
Due to the non-convex form of the optimization problem, its optimal solution is 2-D exhaustive search (TDS) algorithm over BS and relay power variables but it is a highly computationally complex approach.
It is shown that our EE function is a quasi-concave function with respect to each power variable, therefore we propose a method to limit the range of power variables and also PB-ODS method which uses a power bisection algorithm (PBA) with respect to one of the power variables following an exhaustive search with respect to the other power variable.
It is also suggested to use the PBA with respect to two power variables alternatively as the sub-optimal alternative optimization (AOP) algorithm.
This method has a slight but acceptable degradation in performance while its computational complexity has dramatically decreased.

The rest of the paper is organized as follows. The system model for a massive MIMO relay cellular system and some signaling parameters are introduced in section \ref{System model}. 
Section \ref{ProbFormulation} describes the EE optimization problem formulation. 
Using some features of the objective function, we propose an iterative algorithm for solving the problem in section \ref{PrpMeth}. The simulation results are presented in section \ref{SimResults} and finally,  section \ref{SecConclusion} concludes the paper.

\textit{Notations}- Throughout , matrices are written in capital bold letters such as $\mathbf{A}$, vectors are in bold lower case such as $\mathbf{a}$, and scalars as plain letter such as $A$ or $a$.
$\mathbf{A}^T$ denotes the transpose of $\mathbf{A}$, $\mathbf{A}^h$ is the conjugate transpose of $\mathbf{A}$, $\mathbf{A}^{-1}$ represents the inverse of $\mathbf{A}$, $\mathbf{A}^{\dagger}$ corresponds to the pseudo- inverse of $\mathbf{A}$, and $tr(\mathbf{A})$ denotes its trace.
$\mathbb{E}(.)$ represents the expectation operator.

\section{System model}\label{System model} 
\begin{figure}[H]
\begin{center}
\includegraphics[scale=0.3]{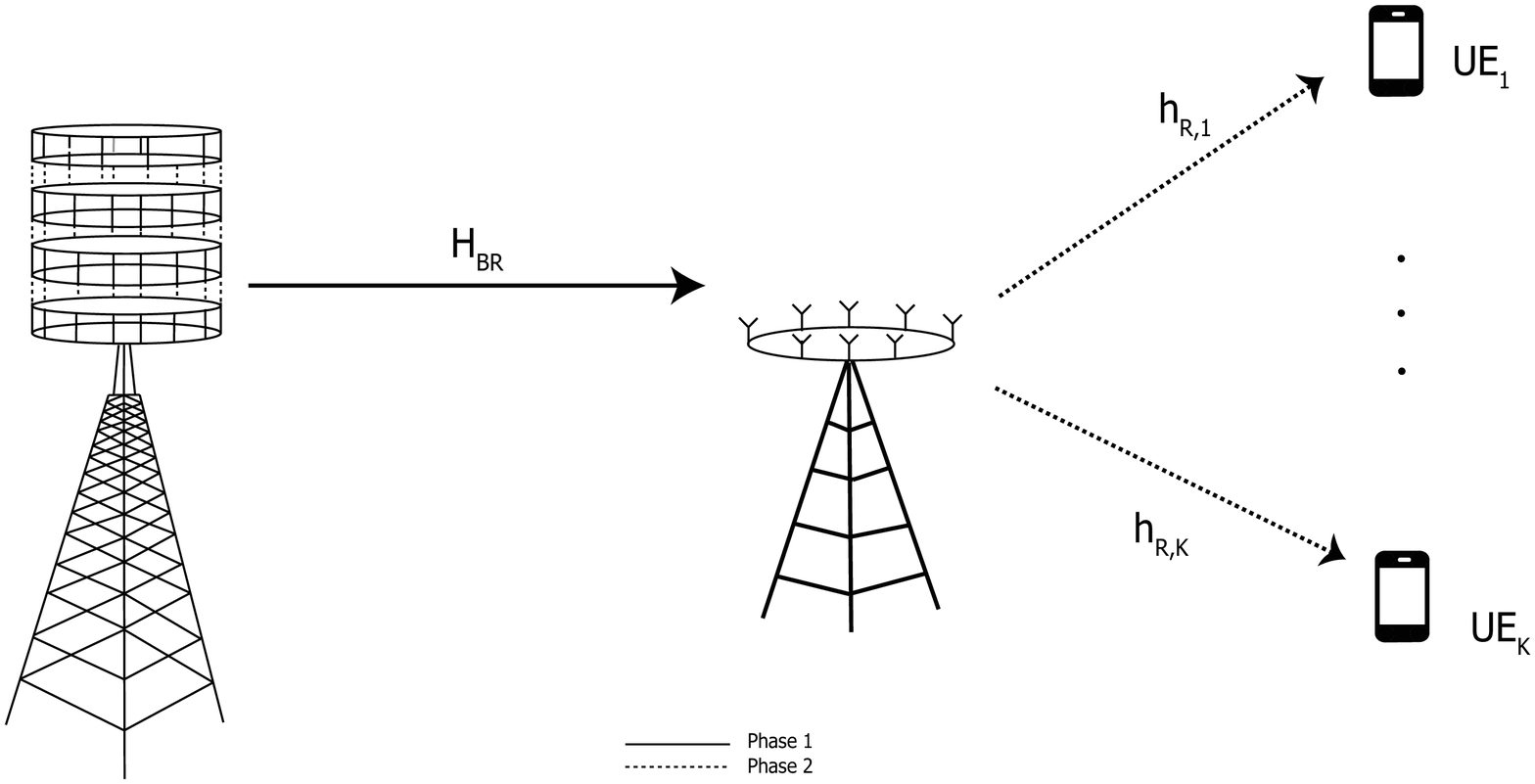}
\caption{Downlink transmission through a massive MIMO relay in a single cell system.}
\label{dl}
\end{center}
\end{figure}

Consider a single cell containing a BS at its center with massive number of antennas $N_{b}$, a MIMO AF relay with $N_{r}$ antennas, and $K$ single-antenna users that are randomly distributed in a cell according to Fig. \ref{dl}. The number of relay antennas is lower than the number of BS antennas $(N_{b} = \alpha N_{r}~, ~\alpha \geq 1)$. 
The relay operates in half-duplex mode which means that data transmission from the BS to its corresponding users is performed in two consecutive time slots. 
Achievable data rate and power consumption model are presented in the following.

\subsection{Achiveable Sum Rate}
According to Fig. \ref{dl}, in the first time slot the BS sends data to the relay. The data at the BS is denoted as $\mathbf{\tilde{s}}_{b}\in \mathcal{C}^{K\times 1}$ such that $\mathbb{E}(\mathbf{\tilde{s}}_{b}\mathbf{\tilde{s}}_{b}^{h}) =\mathbf{I}_{K}$, so the signal that is sent from the BS can be expressed as
\begin{equation}
\mathbf{x}_{b} = \mathbf{F}_{b}\mathbf{s}_{b},
\end{equation}
where $\mathbf{F}_{b} \in \mathcal{C}^{N_{b}\times N_{r}}$ is the precoding matrix at the BS. 
The matrix $\mathbf{F}_{b}$ for ZF proccessing is equal to $\alpha_{b} \mathbf{H}_{BR}^{\dagger}$ where $\mathbf{H}_{BR} = \sqrt{d_{BR}}{\mathbf{G}}_{BR}$ is the channel matrix between the BS and relay. $d_{BR}$ and $\mathbf{G}_{BR}$ are the large-scale log-normal fading coefficient and the small-scale zero mean unit variance complex normal fading matrix, respectively. 
$\mathbf{s}_{b}$ is formed by adding a $(N_r-K) \times 1$ dimension zero vector to $\tilde{\mathbf{s}}_{b}$, so we have $\mathbf{s}_{b}=\left(\begin{smallmatrix} \tilde{\mathbf{s}}_{b} \\ 0_{(N_r-K) \times 1} \end{smallmatrix}\right)$
such that
$\mathbb{E}{(\mathbf{s}_{b}{\mathbf{s}_{b}}^{h})} = \left(\begin{smallmatrix}\mathbf{I}_{K}&0\\0&0\end{smallmatrix}\right) $.
It is also assumed that full channel state information (CSI) is known at the BS.
The power normalization factor ($\alpha_b$) is derived from the fact that the instantaneous transmission power of the BS per timeslot is  equal to $P_b$ and can be expressed as
\begin{align} \label{eq1}
\begin{split}
tr[\mathbb{E}(\mathbf{x}_{b}\mathbf{x}_{b}^{h})] = P_{b} & \Rightarrow tr[(\alpha_{b} \mathbf{H}_{BR}^{\dagger})\mathbb{E}(\mathbf{s}_{b}\mathbf{s}_{b}^{h})(\alpha_{b}\mathbf{H}_{BR}^{\dagger})^{h}] = P_{b} \\
& \alpha_{b}^{2} tr[\tilde{\mathbf{H}}_{BR}^{\dagger}(\tilde{\mathbf{H}}_{BR}^{\dagger})^{h}] = P_b \\
& \alpha_{b} = \sqrt{\frac{P_{b}}{tr(\tilde{\mathbf{H}}_{BR}^{\dagger}(\tilde{\mathbf{H}}_{BR}^{\dagger})^{h})}}.
\end{split}
\end{align}
Since the $\mathbf{s}_{b}$ vector contains $(N_{r}-K)$ zero elements, the matrix $\tilde{\mathbf{H}}_{BR}^{\dagger}$ is derived by selecting the first $K$ columns of the matrix $\mathbf{H}_{BR}^{\dagger}$.
The normalization factor is calculated by considering the vector $\tilde{\mathbf{s}}_{b}$ and the matrix $\tilde{\mathbf{H}}_{BR}^{\dagger}$.
The transmitted data from the BS is received at the relay after passing through the BS-to-relay fading channel. So the signal vector received at the relay can be expressed as
\begin{align}
\mathbf{y}_{r} = \mathbf{H}_{BR}\mathbf{F}_{b}\mathbf{s}_{b} + \mathbf{z}_{r}
=  \alpha_{b}\mathbf{s}_{b} + \mathbf{z}_{r}.
\label{eq_yR1_dl}
\end{align}
It is assumed that the distance between the BS antennas and relay antennas is much greater than the distance between the relay antennas with each other (and also the distance between the BS antennas from each other). Also, $\mathbf{z}_{r}\sim \mathcal{CN}(0,\sigma^{2}_{r}\mathbf{I}_{N_{r}})$ is the additive white Gaussian (AWG) noise. 
The relay multiplies $\mathbf{y}_{r}$ by the beamforming matrix $\mathbf{W}_{r} \in \mathcal{C}^{N_{r}\times K}$ for amplification and ZF processing. 
Therefore, the data sent from the relay is equal to $\mathbf{x}_{r} = \mathbf{W}_{r}\tilde{\mathbf{y}}_{r}$
where the vector $\tilde{\mathbf{y}}_r$ is obtained by choosing the first $K$ elements of vector $\mathbf{y}_r$. 
The processing matrix is equal to
$\mathbf{W}_{r}= \alpha_{r} \mathbf{H}_{RU}^{\dagger}$,
where $\mathbf{H}_{RU}$ is the channel matrix between the relay and users and $\mathbf{H}_{RU} = \mathbf{G}_{RU}\mathbf{D}^{1/2}$. 
$\mathbf{D}$ is a diagonal matrix containing large-scale log-normal fading coefficients and $\mathbf{G}_{RU}$ is a small-scale zero mean unit variance complex normal fading channel matrix.
The signal sent from the relay is equal to
\begin{equation} \label{eq1}
\begin{split}
\mathbf{x}_{r} & = \mathbf{W}_{r}\tilde{\mathbf{y}}_{r} \\
 & = \alpha_{b}\alpha_{r}\mathbf{H}_{RU}^{\dagger}\tilde{\mathbf{s}}_{b} +  \alpha_{r} \mathbf{H}_{RU}^{\dagger}\tilde{\mathbf{z}}_{r},
\end{split}
\end{equation}
where the vector $\tilde{\mathbf{z}}_r$ is obtained by choosing the first $K$ elements of vector $\mathbf{z}_r$. 
The normalization factor $\alpha_{r}$ is chosen such that 
$tr[\mathbb{E}(\mathbf{x}_{r}\mathbf{x}_{r}^{h})] = P_r$
and can be expressed as
\begin{equation}
\alpha_{r} = \sqrt{\frac{P_{r}}{(\alpha_{b}^{2}+\sigma^{2}_{r})tr(\mathbf{H}_{RU}\mathbf{H}_{RU}^{h})^{-1}}}.
\label{eq_alpha22}
\end{equation}

After all, the received signal for the $k$th user is
\begin{equation}
{y}_{u,k} = \alpha_{b}\alpha_{r}\tilde{{s}}_{b,k} + \alpha_{r}\tilde{{z}}_{r,k} + {z}_{u,k},
\end{equation}
where ${z}_{u,k}$ is zero mean, $\sigma^{2}_{u}$ variance AWG noise at users' antennas. 
Therefore, the instantaneous received signal to interference plus noise ratio (SINR) of the $k$th user is
\begin{equation}\label{eq_gamaDL_ZF}
\gamma_{u,k}=
\frac{\alpha_{b}^{2}\alpha_{r}^{2}}{\alpha_{r}^{2}\sigma_{r}^{2} + \sigma_{u}^{2}}.
\end{equation}
Note that the equation \eqref{eq_gamaDL_ZF} holds if users know the values of $\alpha_{b}$ and $\alpha_{r}$. 
Finally, the overall sum-rate for the downlink transmission can be calculated as follows
\begin{align}
\label{eq_KKH}
R_{sum} &= \sum_{k=1}^{K} R_k \\
\label{EqAch}
R_k &= \frac{W}{2}log_2(1+\gamma_{u,k}),
\end{align}
where the factor $1/2$ is due to the half-duplex relay transmission mode and $W$ is the transmission bandwidth.

\subsection{Power Consumption}
Total power consumption model can be described as,
\begin{equation}
P_{total}= \zeta_b P_b + \zeta_r P_r + P_{sys} ,
\label{P_total}
\end{equation}
where $\zeta_b$ and $\zeta_r$ are the BS and relay amplifier efficiencies, respectively. 
$P_{sys}$ is the system power including power consumption of the RF circuits, digital processors, coding, etc. 
This power can be divided to the following parts:

\begin{itemize}
\item RF circuit power consumption:
The power consumption of the downlink transmission consists of two parts:
BS-to-relay and relay-to-user circuit power consumption denoted by 
$P_{c1} = N_{b}P_{tx} + N_{r}P_{rx}$ and $P_{c2} = N_{r}P_{tx} + KP_{rx}$, respectively.
$P_{tx}$ and $P_{rx}$ are RF circuit power consumption for each antenna in the transmitter and receiver, respectively. 
Due to the half-duplex relay transmission mode, the total RF power consumption can be expressed as
\begin{equation}
P_{c} = \frac{1}{2}(P_{c1})+\frac{1}{2}(P_{c2}).
\label{PCdl}
\end{equation}
\item Coding/decoding power:
Similarly, the power consumption for coding/decoding procedure can be expressed as 
\begin{equation}\label{P_cd_dl}
P_{cd} = \frac{K}{2} (P_{cod}+ P_{dec}),
\end{equation}
where $P_{cod}$ and $P_{dec}$ correspond to coding and decoding power consumptions for each user, respectively.
\item Precoding power consumption:
Similar to \cite{hoydis2013massive}, the power consumption associated with the precoding procedure at the BS and relay are denoted by $P_{pc,BS}$ and $P_{pc,R}$, respectively and can be expressed as
\begin{equation}
P_{pc,BS} = \frac{3N_b^{2}N_r+2N_bN_r}{2LT} + \frac{N_r^{3}}{3LT},
\end{equation}
\begin{equation}
P_{pc,R} =\frac{3N_r^{2}K + 2N_rK}{2LT} + \frac{N_r^{3}}{3LT},
\end{equation}
 where $T$ is the coherence time of the channel.
$L$ is BS operation capability per joule.
Therefore the total precoding power consumption is 
$P_{pc} = 1/2(P_{pc,BS} + P_{pc,R})$.
So the total system power consumption can be calculated from the following equation
\begin{equation}\label{p_totall}
P_{sys}= P_{c} + P_{cd}+ P_{pc} + P_{0},
\end{equation}
where $P_{0}$ is a constant power consumption related to backhaul traffic and cooling.
\end{itemize}

\section{Problem Formulation}\label{ProbFormulation}
According to the sum-rate equation \eqref{eq_KKH} and the power consumption model \eqref{p_totall} , the EE of the downlink transmission can be expressed as
\begin{equation}
\eta_{EE} = \frac{R_{sum}}{P_{total}} = \frac{\frac{W}{2} \sum_{k=1}^{K}log_2(1 + \gamma_{u,k})}{ \zeta_b P_b + \zeta_r P_r + P_{sys} }.
\label{eq_ee}
\end{equation}

The EE value represents the number of transmitted bits for the consumption of one joule of energy and our goal is to find optimal power allocation for BS and relay in order to maximize the EE function  \eqref{eq_ee} considering the minimum and maximum transmission power constraints of the BS and relay alongside with the QoS constraints of users.
For each user, its QoS constraint is represented by its minimum achievable data rate, $R_{min,k}$.
Therefore the EE optimization problem can be formulated as
\begin{subequations}
\begin{alignat}{3}
\label{eq_maxEE_UL}
&\max_{P_b,P_r} ~&&\frac{\frac{W}{2} \sum_{k=1}^{K}log_2(1 + \gamma_{u,k})}{ \zeta_b P_b + \zeta_r P_r + P_{sys} } \\
\label{EqOpt12}
&s.t. && 0 \leq P_b \leq P_{b_{max}}, \\
\label{EqOpt13}
& && 0 \leq P_{r} \leq P_{r_{max}},  \\
\label{EqOpt14}
& && R_k \geq R_{min,k},~k=1,...,K
\end{alignat}
\label{eq_maxEE_UL}
\end{subequations}
where $P_{b_{max}}$ and $P_{r_{max}}$ correspond to the maximum transmission powers of the BS and relay, respectively.
Constraints \eqref{EqOpt12} and \eqref{EqOpt13} indicate the minimum and maximum transmission powers of the BS ans relay, respectively.
Constraint \eqref{EqOpt14} indicates the QoS constraints of users.

\section{Proposed Method}\label{PrpMeth}
By solving the EE optimization problem in \eqref{eq_maxEE_UL}, the optimal values of BS and relay transmission powers can be found. 
The ideal but extremely complex approach is the exhaustive search method. 
This optimal method searches among all possible values of transmission powers but it may not be a good practical solution due to its very high complexity.
Another solution is to optimize the relay and BS power according to behavior of the EE function. 
The structure of our EE function is similar to the EE function of \cite{huang2013energy}. 
It is stated in \cite{huang2013energy} that their EE function is a quasi-concave function.
Therefore, it is reasonable to conclude that our EE function is also a quasi-concave function relative to the relay or BS power variables. 
Simulation results also confirm this conclusion.
Using this property, the optimum value of the EE function can be found using a power bisection algorithm.
So we propose the PBA algorithm for relay or BS power allocation based on the quasi-concavity property of the EE function in order to maximize the EE function.

\begin{algorithm}\label{BPA for relay Power}
 \begin{algorithmic}[1]
  \caption{PBA for relay Power}
 \Require $\rm{P_{r_{max}}~,P_{b}^{(0)},\epsilon_r>0,\kappa,P_{r_{min}}}$
 \Ensure $\rm{\eta_{EE}^{*},~P_{r}^{*}}$
 \State $\rm{initialize~P^{(1)}=P_{r_{min}}}$
 \State $\rm{calculate~~\eta_{EE}~(for~P_r =P^{(1)}~\&~P_{b}^{(0)})~from~eq~\eqref{eq_ee}}$
 \If{$\rm{0\geq \frac{\partial \eta_{EE}}{\partial P_r}}$}
 \State $\rm{P_{r}^{*}~\longleftarrow~P^{(1)}}$
 \State $\rm{\eta_{EE}~\longleftarrow~\eta_{EE}~(for~P_r =P^{(1)})}$
\Break
 \Else
 \While{$\rm{~0<~\frac{\partial \eta_{EE}}{\partial P_r}~~\&\&~~P^{(1)}<P_{r_{max}}}$}
 \State $\rm{P^{(2)}~\longleftarrow~P^{(1)}}$
 \State $\rm{P^{(1)}~\longleftarrow min(\kappa P^{(1)},P_{r_{max}}),~\kappa>1}$
 \State $\rm{calculate~\eta_{EE}^{1}~for~P_r=P^{(1)},~P_b^{(0)}~from~\eqref{eq_ee}}$
 \EndWhile
  \If{$\rm{0 \leq \frac{\partial \eta_{EE}}{\partial P_r}}$}
 \State $\rm{P_{r}^{*}~\longleftarrow~P^{(1)}}$
 \State $\rm{\eta_{EE}~\longleftarrow~\eta_{EE}~(for~P_r =P^{(1)})}$
\Break
  \EndIf
 \EndIf
  \State $\rm{calculate~\eta_{EE}^{2}~for~P_r=P^{(2)},P_b^{(0)}~from~\eqref{eq_ee}}$
 \While{$\rm{\vert\eta_{EE}^{2}-\eta_{EE}^{1}\vert>\epsilon~~}$}
  \State $\rm{\overline{P}~\longleftarrow~\frac{P^{(1)} + P^{(2)}}{2}}$
    \State $\rm{calculate~\eta_{EE}~for~P_r=\overline{P},P_b^{(0)}~from~\eqref{eq_ee}}$
    \If{$\rm{0\leq \frac{\partial \eta_{EE}}{\partial P_r}}$}
     \State $\rm{P^{(2)}~\longleftarrow~\overline{P}}$
           \State $\rm{ \eta_{EE}~\longleftarrow~\eta_{EE}^{2}}$
           \Else
     \State $\rm{P^{(1)}~\longleftarrow~\overline{P}}$
           \State $\rm{\eta_{EE}~\longleftarrow~\eta_{EE}^{1}}$
\EndIf
\EndWhile
\State $\rm{Output~~:~~P_{r}^{*}=\overline{P}~~,~~\eta_{EE}^{*}=\eta_{EE}}$
 \end{algorithmic}
\end{algorithm}

Unfortunately the EE function is not a quasi-concave function of both BS and relay power simultaneously\cite{huang2013energy}. 
So the PBA should be used only for one power variable (power dimension) at each time. 
So it is proposed to apply the PBA in one dimension along with the 1-D exhaustive search in the other dimension which is called the 1-D Search (ODS) algorithm.
Therefore there can be two types of PB-ODS (PBR-ODSB and PBB-ODSR) algorithms depending on the power variable which PBA is applied to. 
PBR-ODSB (PBB-ODSR) correspond to the algorithm where PBA is applied to relay (BS) power and ODS is applied to BS (realy) power.

\subsubsection{Relay power optimization}
\label{SecRelayPow}
When the BS power is assumed to be constant in equation \eqref{eq_maxEE_UL} the relay power optimization can be defined as
\begin{align}
&\max_{P_r}\frac{\frac{W}{2} \sum_{k=1}^{K}log_2(1 + \gamma_{u,k})}{ \zeta_b P_b + \zeta_r P_r + P_{sys} }\nonumber\\
&s.t.~~~0 \leq P_r \leq P_{r_{max}},\nonumber\\
&~~~~~~R_k \geq R_{min,k},~~~~k=1,...,K.
\label{eq_maxEE_SubRelay_Dl}
\end{align}
Since the objective function is quasi-concave with respect to the relay power, the PBA can be used to find its optimum value as described in algorithm 1 where
$\kappa$ is the convergence step and $\rm{P_{r_{min}}}$ is the lowest feasible relay transmission power which is used for faster convergence of the algorithm by limiting the relay transmission power variable.
Using the QoS inequality constraints of each user and some mathematical manipulations, $\rm{P_{r_{min}}}$ can be expressed as
\begin{equation}
P_{r_{min}} = \max_{k=1,...,K} \frac{(2^{2R_{min,k}}-1) M \sigma^2_u}
{\alpha_b^2 - (2^{2R_{min,k}}-1) \sigma^2_u}
\end{equation}
where $M = (\alpha_{b}^{2}+\sigma^{2}_{r})tr(\mathbf{H}_{RU}\mathbf{H}_{RU}^{h})^{-1}$.

\subsubsection{BS power optimization}
\label{SecBSPow}
When the relay power is assumed to be constant in equation \eqref{eq_maxEE_UL} the BS power optimization sub-problem can be expressed as
\begin{align}
&\max_{P_b}\frac{\frac{W}{2} \sum_{k=1}^{K}log_2(1 + \gamma_{u,k})}{ \zeta_b P_b + \zeta_r P_r + P_{sys} }\nonumber\\
&s.t.~~~0 \leq P_{b} \leq P_{b_{max}},\nonumber\\
&~~~~~~R_k \geq R_{min,k},~~~~k=1,...,K.
\label{eq_maxEE_SubBS_Dl}
\end{align}
This sub-problem is similar to the problem of optimal relay power allocation. 
The EE function in this case is also quasi-concave and therefore the PBA for BS power variable can be used to find the optimal solution.
In this case, Algorithm 1 can be used by replacing index $b$ by index $r$ and vice-versa.
Using the same approach of previous part, $\rm{P_{b_{min}}}$ can be expressed as
\begin{equation}
P_{b_{min}} = \max_{k=1,...,K} \frac{(2^{2R_{min,k}}-1) M (\alpha^2_r+\sigma_r^2+\alpha^2_u)}
{\alpha^2_r}.
\end{equation}

\begin{algorithm}\label{AOP Algorithm}
\begin{algorithmic}[1]
\caption{AOP Optimization for~relay~\&~BS~Power}
\Require $\rm{P_{r_{min}},P_{b_{min}},~P_{r_{max}},P_{b_{max}},(\epsilon_b,\epsilon_r)>0,\kappa}$
\Ensure $\rm{\eta_{EE}^{*},~P_{r}^{*},~P_{b}^{*}}$
\State $\rm{initialize~P_{r}^{(0)},~P_{b}^{(0)},~l = 0}$
\State $\rm{calculate~\eta_{EE}^{(l)}~from~\eqref{eq_ee}~for~P_{r}^{0}~,P_{b}^{0}}$
\Repeat
\State $\rm{calculate~P_{r}^{*}~for~fixed~P_{b}^{(l)}~from~\eqref{eq_maxEE_SubRelay_Dl}\rightarrow~P_{r}^{(l+1)}}$
\State $\rm{update~P_{b_{min}}^{(l+1)}}$ as decribed in Section \ref{PrpMeth} Part 1
\State $\rm{calculate~P_{b}~for~fixed~P_{r}^{(l+1)}~from~\eqref{eq_maxEE_SubBS_Dl}\rightarrow~P_{b}^{(l+1)}}$
\State $\rm{update~P_{r_{min}}^{(l+1)}}$ as decribed in Section \ref{PrpMeth} Part 2 
\State $\rm{update~l = l+1}$
\State $\rm{calculate~\eta_{EE}^{(l)}~from~\eqref{eq_ee}~for~P_{r}^{l}~,P_{b}^{l}}$
\Until $\rm{\vert\eta_{EE}^{(l)}-\eta_{EE}^{(l-1)}\vert\leq\epsilon~~}$
\State $\rm{P_{r}^{*}=P_{r}^{(l)}, P_{b}^{*}=P_{b}^{(l)}~and~\eta_{EE}^{*} = \eta_{EE}^{(l)}}$
\end{algorithmic}
\end{algorithm}

The PB-ODS algorithm can mitigate the complexity of the solution compared to exhaustive search in both dimensions but 
the algorithm can still be considered computationally complex since the search for one of optimal power values is the exhaustive search.
For further complexity reduction, it is proposed to use the PBA algorithm alternatively for both dimensions until the convergence of the algorithm which is denoted by AOP algorithm as described in algorithm 2.
As it can be seen, at each iteration the optimum value for one of the power variables is computed using the PBA while keeping the other power variable fixed and then this process is repeated for the other power variable.
According to this algorithm, the relay and BS power in the first part of the algorithm are initialized and the corresponding EE is calculated. 
In each sub problem, the problem of optimal power is solved individually for each power variable according to the Algorithm 1 and after the optimization of both power variables, the convergence condition is checked as the stopping criteria of the algorithm. 
Finally, when the EE value converges, the BS and relay transmission powers are determined from the sub-optimal value of their corresponding problem and the near optimal EE value of the system is also computed from the power variables.

\section{Simulation Results}\label{SimResults}
In this section the simulation results are presented and discussed.
The simulation parameters of the system model are listed at table \ref{alg_DlParameter}.
According to the table, users are uniformly distributed at the edge of a circular cell (800-900 meters from the center of the cell) in a 60-degree sector.
Accoring to the Long Term Evolution (LTE) standard , the bandwidth of each sub-channel is set to be 180 KHz.
The QoS constraint of all users are considered equal to $1 bps/Hz$.

\begin{table}
\caption{Simulation parameters}
\label{alg_DlParameter}
\centering
\begin{tabular}{| p{1.7cm} | p{1.5cm} | p{1.7cm} | p{1cm} |}
\hline \hline
Parameter & Value & Parameter & Value\\
\hline
Cell radius & $900$~m &$K$ & $10$\\   
User distance & $800-900$~m &Relay distance & $425$~m \\
$\eta$ & $3.5$ & $P_0$ & $27$~dBm\\   
$P_{tx}$ & $24$~dBm &$P_{cod}/P_{dec}$ & $29$~dBm\\
$P_{rx}$ & $19$~dBm & $T_c$ & $32$~msec\\
$N_0$ & $-100$~dBm & $W$ & $180$~KHz\\
$P_{b_{max}}$ & $29$~dBm & $R_{min}$ &$1$~bps/Hz\\
$P_{r_{max}}$ &  $23$~dBm & $\alpha$ & $2$\\   
$P_{syn}$ & $28$~dBm & $\zeta_b,\zeta_r$ & $0.3$\\ 
$L$ &$10^9$~op/J & $\kappa$ & $1.1$\\  
\hline
\end{tabular}
\end{table}

The following optimization methods are investigated.
\begin{itemize}
\item Exhaustive search in two dimensions (TDS)
\item PBA for relay power and exhaustive search for BS power (PBR-ODSB)
\item PBA for BS power and exhaustive search for relay power (PBB-ODSR)
\item AOP in two dimensions (AOP)
\end{itemize}
\begin{figure}
\begin{center}
\includegraphics[scale=0.6]{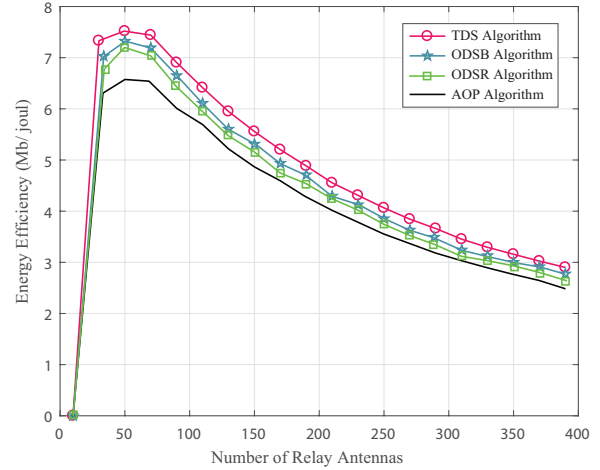}
\caption{Comparison of different algorithms on energy efficiency.}
\label{T4N6_3_4_UL}
\end{center}
 \end{figure}

Fig. \ref{T4N6_3_4_UL} shows the EE of the proposed system model for different power allocation algorithms. 
Also, the performance of the PB-ODS algorithm approaches the performance of the exhaustive search method. 
It can be concluded that the use of the PBA algorithm in one dimension will be suitable for finding the optimal value of the power. 
PB-ODS algorithm still uses an exhaustive search method in one of two dimensions that has a high complexity. 
Although the AOP method achieves lower performance than other algorithms, it is extremely less complicated than other algorithms, which makes it suitable for the cases that the processing speed of the algorithms is important.
It can also be seen that the performance of PBR-ODSB is better that PBB-ODSR.
In other words if the system has the process ability to use exhaustive search in one of the dimensions, it is better to use exhaustive search for relay power and PBA for BS power.
%

\begin{figure}
\begin{center}
\includegraphics[scale=0.6]{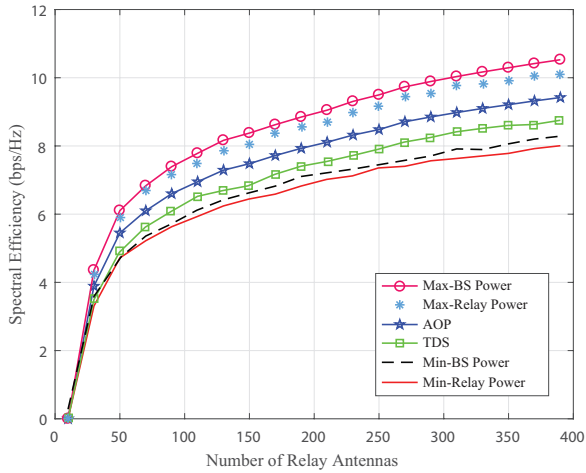}
\caption{The spectral efficiency of the downlink transmission for different power allocation modes (maximum power, minimum power, optimal power and sub-optimal power).}
\label{T4N6_4_2_3_UL}
\end{center}
 \end{figure}

Fig. \ref{T4N6_4_2_3_UL} represents the SE of the system in different modes as they are described in table \ref{Diff_Power}. 
It can be seen that in the case of using maximum transmission power for BS or relay, more SE is achieved for the system. 
However, the aim of the optimization problem is to search for the BS and relay transmission powers in order to maximize the EE of the system while the SE is hold above a specified threshold.
\begin{table}[ht]
\caption{Different power allocation modes}
\label{Diff_Power}
\centering
\begin{tabular}{c|c}
\hline \hline
Description & Name\\
\hline
$P_b = P_{b_{max}}~,~P_r$~from~PBA & Max~BS \\   
$P_b = P_{b_{min}}~,~P_r$~from~PBA & Min~BS \\
$P_r = P_{r_{max}}~,~P_b$~from~PBA & Max~relay \\
$P_r = P_{r_{min}}~,~P_b$~from~PBA & Min~relay \\
$P_r~\&~P_b$~from~PBA & AOP\\ 
$P_r~\&~P_b$~from~Full~Search & TDS\\ 
\hline 
\end{tabular}
\end{table}

\begin{figure}
\centering
\includegraphics[scale = 0.6]{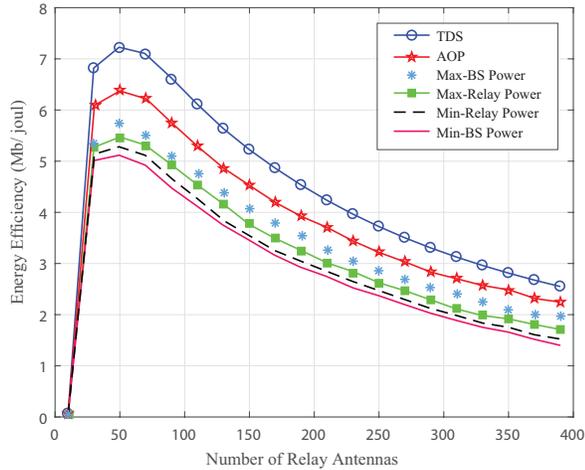}
\caption{Energy efficiency of downlink transmission for different power allocation modes.}
\label{T4N6_4_4_UL}
\centering
\end{figure}

Fig. \ref{T4N6_4_4_UL} shows the EE of the system for different power allocation modes of table \ref{Diff_Power}. 
Due to the usage of ZF pre-coding in the relay and the BS, there is no interference among users. Therefore, it can be seen in the Fig. \ref{T4N6_4_2_3_UL} that increasing the transmission power of relay results in increasing the SE. 
However, it can be seen that the EE of the system with maximum power allocation of the relay and BS is less than optimal and sub-optimal approaches. 
It should be again noted that the aim of this research is to maximize the EE value which may result in SE performance degradation of the system. 

\vspace{-1.5mm}
\section{Conclusion }\label{SecConclusion}
In this paper the EE of a massive MIMO single-cell relay system in downlink transmission is investigated. 
The relay is also assumed to be a massive MIMO antenna which is a less investigated assumption in the literature.
The goal is to serve cell-edge users that face high pathloss.
The problem is to maximize the EE through optimal power allocation of relay and BS. 
The optimal solution can be achieved using the exhaustive search algorithm which has a very high computational complexity.
Considering the quasi-concave behavior of the EE function with respect to each one of power variables, we proposed to use PBA in order to find the optimal value of the EE function with respect to each power variable.
Therefor, we suggest to use the PBA for one of power variables following the exhaustive search method for the other power variable which is called the PB-ODS algorithm.
The simulation results show that the performance of the PB-ODS algorithm approaches the performance of exhaustive search algorithm with less amount of computational complexity.
In order to further reduce the complexity, it is suggested to use the PBA for both power variables alternatively in an iterative manner which is called the AOP algorithm.
Simulation results shows that performance of the AOP algorithm is slightly lower than the high complex optimal exhaustive algorithm but with a dramatic decrease in computational complexity which makes it suitable for practical scenarios.
Also it is shown that due to the quasi concavity behavior of the EE function, maximum power transmission does not necessarily results in the maximum EE of the system.
\bibliographystyle{IEEEtran}
\bibliography{MyRef}
\end{document}